A manifold of possible physics-laws in a universe where the planck constant and speed of light parameters vary

Roee Amit
*Division of Biology and Division of Chemistry and Chemical Engineering*
*California Institute of Technology*
*Pasadena, CA 91125*

Date: February 14[th], 2008

# Abstract

I assume a universe whereby the speed of light and the planck constant are not constants but instead parameters that vary locally in time-and space. When describing motion, I am able to derive a modified path integral description at the quantum level, which offers a natural extension of quantum mechanics. At the microscopic level, this path integral intuitively describes a physics with many "quantum realities' thus leading to a novel concept of "manifold of physics", which can be looked at as a novel action principle. This paradigm reflects the notion that the observed laws of physics on any given scale are determined by the underlying distribution of the fundamental parameters (i.e Quantum Mechanics is just one point on this manifold), thus leading to many possible physical-law based behaviors. By choosing a Gaussian distribution of the parameters, a quadratic action term appears in the path-integral, which in turns leads to a complex classical action (and by continuation a new description for inertia) at the classical level. In the accompanying manuscript the classical doublet equation of motion is applied to the Newtonian gravitation field, and a MOND-like, dark-energy-like, and pioneer-anomaly-like solutions are derived.

# I. INTRODUCTION

The 20th century witnessed two of the greatest intellectual revolutions in the history of humanity in the shaping and understanding of the theories of relativity and quantum mechanics. In particular, the latter has been a major driver for the technology development that ensued in the second half of the 20th century, which has enabled the high-tech society of today. However, both of these great intellectual efforts are flawed, as it is well known that Einstein's general relativity theory, which describes gravity on macroscopic scales is inconsistent with quantum mechanics and its derivatives (i.e. Standard Model) which describes all other behavior on the microscopic scale. The inconsistencies are not just mathematical that generate "nonsense" solutions, but also philosophical ones that point to the heart of the intellectual exercise we call physics – what is the nature of matter, and what is the nature of motion?

Intuitively, we tend to think of matter as particles that are solid spherical-like objects of definite size, shape, mass, spin, etc. that traverse space-time in a precise deterministic path under the auspices of the laws of physics. However, when quantum mechanics came along this concept was altered in a profound fashion much to the chagrin of physicist like Einstein who could not fathom quantum mechanic's brave, new description of the nature of matter. But, despite all its experimental success, can the reality we consider "quantum-mechanical" be examined from a different observational point-of-view? Can we reach the same phenomenological conclusions without having to say that the particles have wave-like properties? Is it possible to get the same results on our detectors while assuming an altogether different more intuitive characteristic for the nature of motion and matter in our universe?

Consider a very naïve observer without any education in physics. That observer is given the task of measuring the position of a microscopic quantum mechanical particle. Needless to say, this naïve observer will not be able to precisely measure the position of the particle and conclude that the system has some inherent uncertainty. The only conclusion he/she can reach in interpreting his/her data is that either the precise position of the particle is uncertain and cannot be determined, or that there is some sort of inherent randomness to the measuring device or method. Since our observer is a very judicious experimentalist and therefore can rule out the device as the source for the uncertainty, he/she will conclude that his/her observed phenomenon is a result of either an inherent uncertainty in the particle's position, or that the gauging particles used for the measurement (i.e. "photons") travel at varying speeds, thus obtaining a different reading for position every time a measurement is read.

The first interpretation will lead naturally to the theory of quantum mechanics, and therefore I shall not dwell on it further. It is the purpose of this paper to explore the ramifications of the alternative conclusion – namely the speed of the gauging particles (speed of light) is not constant. The assumption of a variable speed of light, is not new and has been used in several theories grouped as *vsl* (varying-speed-of-light theories) [1, 2]. In the standard *vsl* paradigm, the variability of the speed of light is invoked for cosmological purposes to form an alternative scenario to the one presented by inflation, whereby in most cases discussed in the literature the speed of light varies as a slowly decreasing function of time (hence its historical value was much larger, allowing for the isotropy and homogeneity observed in the CMB today). In the context of the present work, the variability of the fundamental constants is to be of a different nature. Namely, the speed of light and planck's constant are not universal constants, but rather local parameters that characterize each and every space-time coordinate in the universe. In that sense, we try not to break with special relativity, by assuming that the laws of special relativity hold for each and every coordinate, and thus there are still no inertial frames and particle motion cannot be greater than the speed of light at each and every coordinate of space-time. A particle traversing space-time can essentially be looked upon as "tunneling" from one coordinate to the next, where each coordinate will have its own characteristic values for what we call the constants. The average values of these parameters over a sufficiently large scale will converge to the constant values we recognize. Therefore, the fabric of the universe will be described as "gas-like" material with average "universe-energy" and "universe-temperature", and the observable characteristics of particle motion will be the result of the interaction of regular matter with this hypothetical "gas-like" media.

I will show that by implementing a varying speed of light constraint on the motion of "classical" particles (i.e. ones not having wave-like characteristics), a generalized path integral formulation will emerge as a mathematical description for the motion. The generalized "physics" that will result from this new path integral will be a direct consequence of the underlying assumed distribution of "universon-particles" – namely the space-time distribution of the speed of light and planck's parameter values. This path-integral can be viewed as a mathematical generalization of the quantum-mechanics wave-like paradigm, in the sense that instead of having just one possible quantum reality characterized by $\hbar$, we now have many quantum realities possible characterized by an underlying distribution of a varying planck parameter. Thus, there will be many types of quantum physics possible (a manifold of possible "physics",) where conventional quantum mechanics will correspond to just one point. As a result, each "physics" (or point on this great manifold) will exhibit different possible motions, and other consequences for observables depending on the underlying distribution of the parameters (note, that for convenience I chose a path integral formulation based around the planck parameter, but it is my belief that a completely complementary derivation can be made using the speed of light parameter as well). I named this theoretical scheme *vsl-mechanics*.

In analyzing the *vsl* path integral, I will first use a delta-function distribution for the planck parameter to re-derive the conventional quantum mechanics path integral. Next, I will assume a completely random distribution for the planck parameter and find heretofore unknown physical behavior. The results of a simple analysis will indicate that a length scale characterized by a random distribution of the planck parameter will make the physical measurement of position impossible, and in essence meaningless for that "physics". In fact, if one assumes that the random distribution emerges as a function of decreasing scale, one can therefore attribute to the length-scale where this happens an interpretation of the smallest possible "length" that can have "physical" or observable meaning. This will touch base with similar results obtained from string-theory, and in particular from the doubly special relativity (DSR) [3-5] model that examine physical ramifications to motion whereby a minimum length scale is assumed

Finally, I will assume a Gaussian distribution for the planck parameter which will be characterized by a new parameter (which in this case will be a global constant). The path integral that will emerge from this model will therefore have a quadratic "classical-action" term in the exponent in addition to the usual term. Furthermore, unlike the traditional quantum mechanical model, this path integral will generate radically different classical equations of motion than Newton's laws. The *vsl* action will be complex, and as a result will bring about a new or *vsl* version for the law of inertia. The immediate consequence of this complex classical action is the emergence of two force laws that must be concomitantly satisfied in order to solve for a particle's motion.

In the accompanying manuscript, the new equations of motion will then be applied to the non-relativistic Newtonian gravitational central potential. One of the two equations that will be derived is identical to Milgrom's MOND scheme [6-8], a highly successful and controversial phenomenological model for the missing mass or dark matter. Moreover, the solutions for the second equations when combined with the first will additionally produce a "dark-energy" like non-local force solution, and a pioneer anomaly-like force at a solar system scale. Thus the *vsl*-equations of motion that describe the action of the gravitational field not only reproduce the conventional Newtonian results, but also provide a candidate model that may resolve all three currently unexplained gravitational anomalies: namely, dark matter, dark energy, and the purported pioneer anomaly. Hence, this new *vsl* path-integral formulation will allow us to derive physical behavior on the very largest of macroscopic scales, which is a direct consequence of ultra-microscopic phenomenon and assumptions.

Finally, I remind the reader that it is the purpose of this paper to ask what would the laws of physics be like in a universe characterized by a variable speed of light and planck "constant" parameters, where the variation is only on the very small microscopic scales. This is a legitimate question as the speed of light was never measured on the atomic scale or below, and until this is done, it is not possible to know if it is truly constant or not.

## II. MEASURING OBSERVABLES – THE CASE OF POSITION

In attempting to describe motion in a universe characterized by variable speed of light and other constants we will motivate the approach from hypothetical observations that may be observed in a laboratory setting. Following such an approach will allow us to gain a more intuitive understanding of what kind of conclusions can be drawn given a constraint such as a variable speed of light.

Indeed, an experimental determination of any property or characteristic of the universe, no matter what the underlying physical reality may be, will always reduce at the laboratory level to the measurement of one or more observable quantities. The physical process of measurement is essentially the collection of particles on a detector in a certain window of time and space, such that an analysis of this "raw" signal will yield a data set, which allows the observer to draw conclusions on the phenomenon studied. The particles collected on the detector get there as a result of one of two processes: either the object being observed emits particles through some sort of internal process (e.g. radioactive decay), or the observer illuminates it with a beam of particles, and subsequently gathering what is emitted in return (e.g. scattering, fluorescence, absorption, etc.) to determine the observational reality. In any way, particles collected on the detector can be said to contain information on two general processes: first the "gauging process" (either induced externally by the observer or internally by the target), and then the "collection process" - the combined effects of which is the observational reality induced by the phenomenon that the experimentalist wishes to study.

The information collected on the detector is therefore an amalgamation of two separate processes. What happens if each of the separate processes is indeterminable such that the measurements obtained on the detector can in fact result from a combination of possibilities that all add up to the same observed reality? In a *vsl* universe this is precisely what is expected, and such indeterminacy must be prevalent. Therefore, the observed physical reality in such a universe must reflect the underlying indeterminacy that generated the observation. To make the point clearer let's consider a classical universe, where everything is deterministic, and precisely known. Any result collected on the detector will be precisely repeatable in a predictable fashion, as each of the information processes producing it will have singular values, which albeit not measured directly are still known.

In a *vsl* universe any gauging and collection process whose timing depends on the speed of the particles will yield varying results. But, a certain result can be made up of an infinite combination of gauging and collection possibilities each adding up to the same value. In such a universe, the observer has no way of knowing the precise nature of the individual processes, only infer the variability by repeated experimentation and the observation that each time a different result is recorded, or through the inherent variability predict new phenomenon (e.g. quantum wave-like) that are not possible in a classical universe.

In order to make the subsequent discussion simpler we will assume that there are two mathematical quantities. First there will be the "observable-functions" which will replace the traditional notion of observables, and will depend on the fashion by which the speed of light (or other constants) vary in space and time. The variability of the speed of light as a function of space and time will be described by a mathematical quantity called the "parameter function". This function may not necessarily be real from a mathematical perspective, as it does not necessarily correspond to a real measurable quantity, it does need to be analytic and continuous. As a result of the above discussion, we postulate two conditions that will characterize motion in a *vsl*-universe:

1) Parameters such as the speed of light, planck's constant, etc. are described via a parameter function (e.g. $\psi(c)$) which is related to the probability to have a given value of that parameter at some coordinate of space-time. This function is not a probability distribution, as we cannot directly measure its value. Therefore, the only constraints that we will place on this function is that it will have to be continuous and analytic. All observables (e.g. position, displacement, velocity, etc.) are described via functions that depend on the underlying parameter functions.

2) Motion of all types is described via the classical action.

The meaning of postulates 1) and 2) implies that according to the variable speed of light hypothesis it is not that the particle's motion is imprecise or non-deterministic, rather it is our observations of it that are such. Basically, the observation process of a particle's motion will consist of the convolution of these two conditions (i.e. the nature of the parameter functions and the classical action), which will result in a novel composite description for the motion. This goes to the heart of the question of what is physical reality. The notion of "a particle's motion" is meaningful only if there is someone or something there to observe or be affected by it. Therefore, the laws of physics should be defined from the point of view of what we can and cannot observe.

In order to clarify the above discussion and to develop the basic mathematical formalism by which we can make physical calculations, we will consider first the case of measuring a particle's position in a *vsl* universe as an example. In the process, we will define the position function, which will be proportional to the speed of light function, and will be subsequently used as the basis for all other physical derivations.

Since a given measurement of position will be obtained from the total time it takes the photons to return to the observer in a particular experiment, the instantaneous measured position will invariably depend on the varying speed of light, assumed to be determined by the parameter function $\psi(c)$. Hence, measuring a target particle's position involves a two-step process: first, releasing the gauging particles towards the target, and second having the target back-scatter them (i.e. absorbing them and re-releasing them in the direction of the observer) stopping the watch of the observer when the returning particles are detected. The time it takes the particles to reach the target is given by $\Delta t_n^f$, which has a time function $\psi(\Delta t_n^f)$ that is proportional to $\psi(c_n^f)$, where subscript $n$ corresponds to the nth measurement, superscript $f$ is for the gauging process, and superscript $b$ for the collection process. Since the displacement is given by the time it takes to the particles to travel to the target ($\Delta t_n^f$), plus the time to return ($\Delta t_n^b$), the distribution of the measured times $\Delta t_n$ has to account for all possible values of $\Delta t_n^f$ and $\Delta t_n^b$ that can sum up to its value, therefore,

$$\phi(\Delta t_n) = \int \psi(\Delta t_n^f) \psi(\Delta t_n - \Delta t_n^f) d\Delta t_n^f \qquad (1)$$

Corresponds to the probability distribution to measure

$$x_n^m = c_0 \Delta t_n \qquad (2)$$

thus generating a measurement of fluctuating or uncertain position. The probability distribution of the measured position, which must be a real mathematical quantity, will therefore be given by the exact same distribution as in eqn. (1), but with a slight modification,

$$\phi\left(\frac{x_n}{c_0}\right) = \int \psi'\left(\frac{x_n^f}{c_0}\right) \psi''\left(\frac{x_n^r}{c_0}\right) dx_x^f \qquad (3)$$

where $\psi'\left(\frac{x_n^f}{c_0}\right)$ and $\psi''\left(\frac{x_n^r}{c_0}\right)$ are proportional to he speed of light function, and correspond to the particle's position function. Whatever the point of view, the end-result of the measurement is that of an uncertain position for the particle. But, in both cases, the measured probability distribution is dependent on a more fundamental quantity we term the observable function (the position function in this case).

The existence of the resultant parameter and the observable functions is the mathematical expression of the first postulate, and in as much it will enter into any expression that describes physical reality. In a sense, it replaces the coordinate system used in the standard classical or quantum universe. That is, rather than having a grid of n dimensional coordinate system to describe a particle's motion, each coordinate representation is now replaced by a parameter function that is constrained by the underlying distribution of the speed of light values. The mathematical properties of the observable function can be constrained only by the measurable physical reality. Since the behavior of observable, in this case the position, is determined by probability distributions as in eqn (1) and (3), the parameter functions can take any form as long as the integral in those equations forms a continuous and real mathematical function. Therefore, we have considerable freedom in defining the mathematical properties of the parameter functions.

### III.  MEASURING DISPLACEMENT AND VELOCITY

The next quantity that we would like to explore is the change in position or displacement of the target relative to the observer, or equivalently the velocity. In other words, we will treat the case of a free moving particle relative to an observer without the presence of external forces. We will follow a similar logic path to the one shown in the above section. In every displacement measurement, or for that matter velocity, we need two readings of the position: the position $x(t_1)$ and the position some time later $x(t_2)$. The displacement vector then becomes,

$$\vec{d} = \vec{x}(t_1) - \vec{x}(t_2) \tag{4}$$

which is just the difference between the two position measurements. In the *vsl* universe the values for the postions $x(t_1)$ and $x(t_2)$ will vary, such that in calculating the displacement function we must account for all possible combinations of $x(t_1)$ and $x(t_2)$ that will result in a given quantity $d$, which means that the displacement function will be given according to (note that this is not the probability distribution to measure the displacememt):

$$\chi(\vec{d}) = \int \psi(\vec{x}(t_1) + \vec{\varepsilon}) \psi(\vec{x}(t_2) + \vec{\varepsilon}) d\vec{\varepsilon} \tag{5}$$

changing variables we get,

$$\chi(\vec{d}) = \int \psi(\vec{\varepsilon}') \psi(\vec{\varepsilon}' - \vec{d}) d\vec{\varepsilon}' \tag{6}$$

where,

$$\vec{\varepsilon}' = \vec{x}(t_1) + \vec{\varepsilon} \tag{7}$$

Expressing eqn. (6) in terms of Fourier transform functions we get,

$$\chi(\vec{d}) = \int \hat{\psi}(\vec{q})\hat{\psi}(-\vec{q})\exp(-i\vec{q}\cdot\vec{d})d\vec{q} \tag{8}$$

Finally, the velcoity function can then be expressed simply as :

$$\chi(\vec{v}) = \int \hat{\psi}(\vec{q})\hat{\psi}(-\vec{q})\exp(-i\vec{q}\cdot\vec{v})d\vec{q} \tag{9}$$

Eqn (10) and (11) correspond to the most general expression for the measured velocity and displacement of a free-moving particle in a universe characterized by a variable speed of light. Note, that in the constant-c limit as $\psi(c) \to \delta(c-c_0)$ we get that the position and the velocity functions approach the classical limit singular value as well, and become $\phi(\vec{x}(t)) \to \delta(\vec{x}(t)-\vec{x}_0(t))$ and $\chi(\vec{v}(t)) \to \delta(\vec{v}(t)-\vec{v}_0(t))$ respectively. Thus, reducing to the standard non-*vsl* coordinate system, where each coordinate on the grid is characterized by a singular value rather than a parameter function, and a given path can be described precisely in a deterministic fashion.

## IV. MOTION UNDER THE INFLUENCE OF EXTERNAL FORCES.

Determining the motion of particles, which move under the influence of external forces will require a similar approach and treatment. The motion of any particle is described mathematically by the path it traverses in space-time. Classically, a singular path is determined by the minimization of a quantity called the action integral, and through the solution of the resultant Euler-Lagrange equations. In the *vsl* universe, there cannot be one minimized path for a given particle, as each position along the path is undetermined. In fact, due to the random nature of the speed of light, repetitive observation of moving particles under the same conditions will result in different observed paths for every measurement. In order to develop a "path" function describing motion in a *vsl* universe, we first divide a given hypothetical path $\vec{P}(x,t)$ to small segments of length $\varepsilon$ :

$$\vec{P}(x,t) = \sum_n \vec{p}(x_{n+1}, x_n, \varepsilon) \tag{10}$$

As such, motion in each segment can be described as a small displacement from position $x_n$ to $x_{n+1}$, whose "displacement-function" will be given according to eqn. (6) of the previous section. We designate this quantity as:

$$\Psi(seg_n) = \int \psi(x_n)\psi(x_{n+1})d\Delta_{n \to n+1} \tag{11}$$

where $\psi(x_n)$ and $\psi(x_{n+1})$ are the position funtions for $x_n$ and $x_{n+1}$ respectively, and $\Delta_{n \to n+1}$ corresponds to all possible segment lengths as follows,

$$\Delta_{n \to n+1} = \vec{x}_{n+1}(c_{n+1}) - \vec{x}_n(c_n) \tag{12}$$

Now since a path is made of many such segments it is only natural, that the path function will be expressed as follows:

$$\Phi(path) = \Psi(seg_0)\Psi(seg_1)\ldots\Psi(seg_n) \tag{13}$$

$$\Phi(path) = \int \psi(x_1)\psi(x_1 - \Delta_{0 \to 1})\psi(x_2)\psi(x_2 - \Delta_{1 \to 2})\ldots\psi(x_{n+1})\psi(x_{n+1} - \Delta_{n \to n+1}) d\Delta_{0 \to 1} d\Delta_{1 \to 2} \ldots d\Delta_{n \to n+1} \tag{14}$$

which expresses the infinite combination of possible values that can be obtained for the different path intervals, and hence corresponds to a description of all possible paths.

The above description is akin to a coordinate by coordinate description of the path, and as a result is hard to work with from a physics perspective. If we want to derive equations of motion, a more illuminating description is needed. To that end, we will make use of the second postulate, that motion is derived from the minimization of the classical action in a guise that accounts for the *vsl* nature of the universe.

In order to derive the *vsl*-path integral, we will rely in the following on Feynman's famous 1948 paper *Space-Time Approach to Non-Relativistic Quantum Mechanics* [9]. Classical mechanics describes the motion of a particle as a path that is generated by the minimization of a quantity called the action integral. When the path itself is broken down to many infinitesimal segments, each segment of motion corresponds to the minimization of its own infinitesimal action. The action integral is a scalar defined as the time integral of the classical Lagrangian function $L(\dot{x}(t), x(t))$

$$S[x(t)] = \int L(\dot{x}(t), x(t)) dt \tag{15}$$

This can also be expressed as a sum of the infinitesimal actions along path segments, in accordance with Feynman's definition:

$$S = \sum_i S(x_{i+1}, x_i) \tag{16}$$

where,

$$S(x_{i+1}, x_i) = Min \int_{t_i}^{t_{i+1}} L(\dot{x}(t), x(t)) dt \tag{17}$$

In the *vsl* universe the above two equations are modified. Since every measurement of position will differ, every infinitesimal action will be different, and as a result minimized accordingly. This means that for every segment

$$S[x_{i+1}(c_{i+1}), x_i(c_i)] = \varepsilon L\left[\frac{x_{i+1}(c_{i+1}) - x_i(c_i)}{\varepsilon}, x_{i+1}(c_i)\right] \quad (18)$$

there are many possible paths that can be taken, and the total path will be some combination of the particular infinitesimal paths taken for each segment. Therefore, eqn. (17) in the *vsl* universe will become,

$$S[x_{i+1}(c_{i+1}), x_i(c_i)] = Min \int_{t_i(c_i)}^{t_{i+1}(c_{i+1})} L(\dot{x}(t), x(t)) dt \quad (19)$$

$$S[x_{i+1}, \Delta_{i \to i+1}] = \varepsilon L\left[\frac{\Delta_{i \to i+1}}{\varepsilon}, x_{i+1}(c_{i+1})\right] \quad (20)$$

Where the $x_i(c_i)$ correspond to a particular reading of position (as determined by the path function) in the ith segment.

In order to determine a proper description for a path in the spirit of the path and parameter functions described above, we need to define a density function that will properly include all possible paths, and whose integration over all the paths will generate the proper "path function" as a generalization of the displacement case. Since there is a whole distribution of possible end points for each path segment, there is, therefore, a corresponding distribution of actions that describe the potential path along such a segment. Thus, we define the path function as follows:

$$\Phi(\Delta_{k \to k+1}) \equiv \int \Psi\{S[x_k(c), x_{k+1}(c)]\} d\Delta_{k \to k+1} \quad (21)$$

$$\Phi(\Delta_{k \to k+1}) \equiv \int \Psi\{S[x_{k+1}(c), \Delta_{k \to k+1}]\} d\Delta_{k \to k+1} \quad (22)$$

where $\Delta_{k \to k+1}$ corresponds to the k[th] path segment, and $\Psi\{S[x_k(c), x_{k+1}(c)]\}$ is the path functional density, such that each action-coordinate of the functional corresponds to a particular speed of light, and whose minimization will induce a different path along $\Delta_{k \to k+1}$. When discussing paths we need to account for the total path, not just for a particular segment. The path function is the same for all segments. However, once a certain path is chosen in segment $\Delta_{k \to k+1}$, the same set of possibilities is now possible for the next segment $\Delta_{k+1 \to k+2}$, and so on. Therefore, for the total path we have a path functional that starts at some origin and ends at some end point as follows:

$$\Phi(path) \equiv \int \Psi\{S[x_0(c), x_1(c)]\}\Psi\{S[x_1(c), x_n(c)]\}\ldots\Psi\{S[x_{n-1}(c), x_n(c)]\}d\Delta_{0\to 1}d\Delta_{1\to 2}\ldots d\Delta_{n-1\to n} \quad (23)$$

The above path functional is the most general expression for the description of a particle's motion in the *vsl* universe, where each segment $\Delta_{k\to k+1}$ is infinitesimal in size. In a sense it corresponds to the mathematical expression of postulates 1 and 2, and can be viewed as the *vsl* action.

It is easy to show that the mathematical expression that was reached in eqn. (14) and the one in eqn. (23) are in fact equivalent (up to some constant), as they describe the same physical reality. If we use the following definition:

$$\Psi'(f(\Delta_{k\to k+1}, x_{k+1})) \equiv \psi(x_{k+1})\psi(x_{k+1}, \Delta_{k\to k+1}) \quad (24)$$

we then see that eqn. (16) becomes:

$$\Phi(path) = \int \Psi'(f(\Delta_{0\to 1}, x_1))\Psi'(f(\Delta_{1\to 2}, x_2))\ldots \Psi'(f(\Delta_{n\to n+1}, x_{n+1}))d\Delta_{0\to 1}d\Delta_{1\to 2}\ldots d\Delta_{n\to n+1} \quad (25)$$

Both the path integrals corresponds to functions that take into account all possible paths that can be measured for a particle moving in a universe with a variable speed of light. Now taking into account eqn. (23), we see that:

$$\Psi\{S[x_{n-1}(c), x_n(c)]\} = \Psi\{S[\Delta_{n\to n+1}, x_n]\} \prec \Psi'(f(\Delta_{n\to n+1}, x_n)) \quad (26)$$

since both $S[\Delta_{n\to n+1}, x_n]$ and $f[\Delta_{n\to n+1}, x_n]$ are analytic functions of $\Delta_{n\to n+1}$ & $x_n$, $f(\Delta_{k\to k+1}, x_{k+1})$ is, therefore, proportional to the classical action, and eqn. (23) is an extension of Feynman's path integral, or in other words the path integral describing motion in *vsl* mechanics.

Finally to put this in a more familiar form, that will also provide a good discussion basis for the rest of the paper, we will rewrite eqn. (23) as follows using the Fourier transform,

$$\Phi(path) \equiv \int \hat{\Psi}\{\alpha_0\}\exp(i\alpha_0 S[x_0(c), x_1(c)])\ldots \hat{\Psi}\{\alpha_{n-1}\}\exp(i\alpha_{n-1}S[x_{n-1}(c), x_n(c)])d\alpha_0\ldots d\alpha_{n-1}d\Delta_{0\to 1}\ldots d\Delta_{n-1\to n} \quad (27)$$

The above is the most general expression describing motion in a *vsl* universe. What we obtained is very similar to Feynman's path integral, and can in fact be looked at as a generalization of the Feynman approach. It is assumed that the integral of eqn. (27) contains all that is needed in order to derive a coherent description of motion in a *vsl*-universe. By loosening the edict that the speed of light has to be constant, we derived a description of motion, which contains quantum mechanics, but also implies strongly to a connection between the speed of light and planck constant parameters. That is, a variable speed of light automatically implies a variable planck constant. This connection between the planck and speed of light parameter means that the expression in eqn. (27) corresponds to a generalization of the quantum mechanical wave-like approach, and as a result generates an additional (equivalent) interpretation to observations in a *vsl*-universe. Instead of discussing measurements as having a variable speed of light nature, one can equally describe nature as having multiple quantum realities, where a quantum reality is defined by a given distribution of the planck parameter. In a sense, eqn (27) contains a manifold of possible physics laws of motion, where each point on the manifold corresponds to a given form $\hat{\Psi}\{\alpha\}$, which in turn will generate a different path integral with its own resultant equations of motion or physics. Indeed, the quantum mechanics path integral emerges simply from the following path-density functional:

$$\hat{\Psi}\{\alpha_k\} = \delta\left(\alpha_k - \frac{1}{\hbar}\right) \tag{28}$$

which can then be looked at as one such point on the manifold of possible "physics".

The *vsl*-formalism developed thus far allows for a logical generalization of quantum mechanics and the physics that it entails. Indeed, according to *vsl*-mechanics the very existence of quantum mechanics implies an entire landscape of other possible sets of physics law of motion. We will dedicate the rest of the manuscript to exploring the "physics" laws that emerge from two particular choices for $\hat{\Psi}\{\alpha\}$, and subsequently derive a more comprehensive theory that will assume "particle-like" characteristic to the fabric of the universe that will constrain somewhat this physics manifold.

## V. ABSOLUTE ZERO OF DISTANCE

As we have seen, the intimate connection between the planck parameter function $\hat{\Psi}\{\alpha_k\}$ and the speed of light parameter function $\psi(c)$ can yield many possible physical realities. Quantum mechanics as given by equations (27) and (28) simply correspond to a one extreme case for the many possible forms for $\hat{\Psi}\{\alpha_k\}$. In this section we will explore another extreme case, whereby each segment can only take one possible value for *c* at a given moment, thereby implying a mixture of an infinite amount of quantum realities. In mathematical terms this can be expressed as follows:

$$\hat{\Psi}\{\alpha_k\} = \exp\left(-i\alpha_k S_k[x_k(c_k), x_1(c_k)]\right) \tag{29}$$

The purpose of this case is to spell out an alternative reality that can be possible in a *vsl* universe, and compare and contrast it with the other extreme case of quantum mechanics. The meaning of the above distribution is in a sense "opposite" to that of quantum mechanics case described in the previous section. Instead of an infinite set of possible speeds of light, we choose a reality that every segment of our path is characterized by one speed of light (of some value), but which will differ from segment to segment. The effects on the quantum reality description are simple. In this case we will have a mixing of an infinite set of possible quantum constants, which will results in a novel physical reality. Inserting the above relationship into eqn (27) for every $\hat{\Psi}\{\alpha_k\}$, and integrating over the $\alpha_k$ we obtain:

$$\Phi(path) \equiv \int \delta\{S[x_0(c), x_1(c)] - S_0[x_0(c_0), x_1(c_0)]\}\ldots\delta\{S[x_{n-1}(c), x_n(c)] - S_{n-1}[x_{n-1}(c_{n-1}), x_n(c_{n-1})]\} d\Delta_{0\to 1}\ldots d\Delta_{n-1\to n} \qquad (30)$$

This path integral means that there will be only one possible path that can be observed in every segment, and this path will minimize the action defined be some arbitrary value of the speed of light $c_k$. In a sense, a classical behavior for each segment – but not really. In the quantum mechanics case the path integral reduced to all possible paths over the entire set of segments, which can be interpreted as a moving wave-front of possible outcomes. In this case there is one possible displacement value for every segment measurement. However, since the motion in every segment is independent of every other segment, the total path itself is still undetermined. That is, if an observer is to carry out consecutive measurements of path length on this scale, he/she will get classical-like values and behavior (i.e. no wave-like phenomenon such as interference), that will vary for each measurement on a scale from 0 to infinity with equal probability, as in every measurement a different average speed of light $c_k$ will define the classical observed behavior, while there will be a mixing of an infinite set of quantum realities thereby canceling each other's individual wave-like characteristics. As a result, the only conclusion an observer within this "universe" will be able to draw is that any measured speed of light related observable is completely indeterminable and therefore has no physical meaning.

To see this, consider the case of displacement measurement: if an observer attempts to measure a distance to an object, the values of successive measurements will vary from 0 to infinity with equal probability. Thus, the whole concept of position as we understand it, will be completely ill-defined in such conditions. However, if we consider a universe whose properties are such that at a given scale:

$$P(x) = const, 0 < x < \infty, d \leq d_{ums}, x \in d_{ums} \qquad (31)$$

where $d_{ums}$ defines the transition into the random quantum reality phase as in this case, and for scales characterized $d \geq d_{ums}$ we will expect the following probability distribution to measure a given particle's position:

$$P(x) = f(x), 0 < x < \infty, d \geq d_{ums}, x \in d \qquad (32)$$

such that, if a distribution such as shown in eqn. (31) exists on some size scale bigger than 0 it will be considered the smallest scale at which any meaningful physical phenomenon can be measured – meaning that at this scale and below measurements for position, displacement, and velocity will be completely random and ill-defined. A distribution such as this will mean that the length scale that this occurs at can be termed in a way "*the absolute zero-of-length*". There will be no meaning to lengths smaller than this size-scale, as any measurement on a scale "smaller" will result in precisely the same random distribution of results, and will therefore be completely indistinguishable from the larger scale. In this case a total arbitrary mixing of quantum realities results in a prediction that the length scale at which this occurs will correspond to the smallest size scale that can be possible in a universe from a physics observational perspective.

By exploring the two extreme cases (i.e. having delta function values for both $\hat{\Psi}\{\alpha_k\}$ and $\psi(c)$), we were then able to examine and speculate on the effects of having a universe with multiple quantum realities. When $\hat{\Psi}\{\alpha_k\}$ is singular we have the full slew of possible quantum wave-like characteristics. As the amount of possible quantum realities increases, it is tempting to speculate that the effect of this on physical behavior is to limit and constrain the wave-like characteristics that is observed on the standard microscopic scale. The present case brings to light the situation whereby an infinite set of quantum realities is allowed thus leading to complete cancellation of the wave-like behavior in the scale in question, thus leading to a seemingly deterministic behavior of particles. Although, such a solution may seem disturbing at first, it carries with it the possibility that a final non-zero scale may exist, such that smaller length scales below would be physically excluded on the grounds of indistinguishability as a result of the physics laws that govern this scale.

Since there is no particular reason as to why one type of solution will be favored over another in the *vsl* context, it would seem reasonable to assume that if quantum mechanics is a physical reality, then so should the "*absolute zero of length*" scale be a physical possibility. In a sense, we require a theory that will have the single quantum reality state as one physical limit and the to '*zero-of-absolute-length*' scale as the other. Such a theory will correspond to an attractive first principle model for *vsl* mechanics that will comprise a manifold of possible physics that span these two extreme cases, and everything else in between. Having a theory that derives a "*zero-of-absolute length*" at ultra-microscopic size scales is attractive to many modern-day physical conundrums, particularly for gravitational singularities that lie at the heart of the conflict between quantum mechanics and general relativity. Indeed, a minimum length result has been obtained by superstring theory [10], and used as part of its solution to the previously mentioned conflict. Moreover, an extension of special relativity that was designed to precisely deal with such a physical reality was constructed recently and termed doubly special relativistic (DSR) [3-5]. This theory attempted to extend the Lorentz symmetry to the planck length scale, which was deemed an irreducible length scale viewed at the same value by all observers irrespective of their speed and scale. Of the many effects that were discovered by this extension – a connection to *vsl* concepts was considered, and in particular a prediction that $\hbar$ was 0 at the planck length scale was obtained in agreement with the above discussion [5].

It is the goal of the next section, and indeed the rest of the paper, to make a first attempt to derive a more comprehensive *vsl* theory that will span the physics manifold, and in the process explore the ramifications of the *vsl*-phenomenon at the macroscopic scales. This theory will be developed from a statistical thermodynamics perspective having the universe itself be described via a "thermal" gas of "quasi-particles" called universons having "universon-energy" and "universon-temperature". It will be assumed that the average universon energy is none other than the planck's constant, and each universon will therefore be characterized by its own speed of light and "universon-energy".

## VI. UNIVERSON STATISTICAL MECHANICS

Thus far, we attempted to describe a universe where the speed of light is not constant, but rather varies in space and time for each coordinate in the universe. This assumption has lead to the derivation of a path integral formalism, from which one can derive the resultant the equations of motion. We concluded that *vsl*-mechanics was a theory that can be viewed as a generalization of quantum mechanics, as the latter is one particular point on a landscape or manifold of possible "physics". Indeed, the derived path integral (eqn. (27)) was too loosely defined in allowing the distribution $\hat{\Psi}\{\alpha_k\}$ to remain undetermined, as was shown by the two extreme cases discussed that lead to quantum mechanics and the special "absolute zero of distance" physics respectively.

It is the purpose of this section to constrain the *vsl*-physics manifold by introducing an additional assumption on the way by which the planck and speed of light parameters vary in time space. That is, the so-called distribution of the speed of light and the planck parameters characterize an "ideal gas" of non-interacting ultra-microscopic particle-like objects called universons, whereby each universon is characterized by its own set of values for the planck, speed of light, and possibly other parameters. Therefore each "particle" in the gas of universons can be described as having a "universon-energy" $\alpha_k$ and occupy associated energy levels $A_k$, such that the following is true:

$$\sum_k A_k = A \tag{33}$$

$$\frac{1}{A}\sum_k \alpha_k A_k = \tilde{\alpha} = \frac{1}{\hbar} \tag{34}$$

Where A is total number of available universon energy states, and $\tilde{\alpha}$ is the average universon energy, which is the reciprocal of the conventional planck constant. We make these assumptions based on our record of large experimental body of data that leads us to believe that $\hbar$ is a constant on all-probed microscopic scales and above. The total number of distinct possible ways $\Omega(A_1, A_2 \ldots)$ of selecting a total of *A* distinct systems in such a way that $A_1$ of them are in state k=1, $A_2$ of them are in states k=2, while satisfying the above constraints is given by simple combinatorial reasoning:

$$\Omega(A_1, A_2 \ldots) = \frac{A!}{A_1! A_2! \ldots} \tag{35}$$

In the limit where the number of particles that make up this ensemble becomes very large $\Omega(A_1, A_2 \ldots)$ grows exponentially as a function of the number of particles. We can therefore set the following:

$$\Omega(\tilde{\alpha}) = \Omega(A_1, A_2, \ldots) \tag{36}$$

where, for convenience we define the universon entropy as follows:

$$\sigma \equiv \ln(\Omega(\tilde{\alpha})) \tag{37}$$

A well known result from statistics then claims (see [11] 6.10) that the probability to find this system in some "universon-energy" state $A_k$, which is characterized by energy $\alpha_k$ is described by the Boltzman distribution and given according to:

$$P_k = \frac{\exp(-\beta \alpha_k)}{\sum_k \exp(-\beta \alpha_k)} \tag{38}$$

where $\beta$ is the universon temperature parameter for this case. Hence, the probability of finding $A$ in one particular state $k$ of energy $\alpha_k$ is given by equation (38). In the limit of a large ensemble, since the number of states is so large, one is interested in measuring the probability that $A$ has energy in a small infinitesimal range given by:

$$\tilde{\alpha} < \alpha < \tilde{\alpha} + \delta\tilde{\alpha} \tag{39}$$

this number is simply given by adding the probabilities for all the states whose energies are found in this range as follows,

$$P(\alpha) = \sum_r P_r \tag{40}$$

since all of these states can be found in a very small energy range, $P_r$ is characterized essentially by the same exponential given by equation (38). Therefore, the probability $P(E)$ is simply given by the number of possible states $P_r$ multiplied by the distribution described by equation (40). Since the number of states was calculated to be $\Omega(\alpha)$ and is given by equations (35) and (36), we have:

$$P(\alpha) = C\Omega(\alpha)\exp(-\beta\alpha) \tag{41}$$

Taking the logarithm of both sides and expanding around $\tilde{\alpha}$ we get,

$$\ln[P(\alpha)] = C\left\{\ln[\Omega(\tilde{\alpha})] + \frac{\partial \ln \Omega}{\partial \alpha}(\Delta\tilde{\alpha}) + \frac{1}{2}\frac{\partial^2 \ln \Omega}{\partial \alpha^2}(\Delta\tilde{\alpha})^2 - \beta(\tilde{\alpha} + \Delta\tilde{\alpha})\right\} \tag{42}$$

adding terms, and using the standard statistical mechanics definition for $\beta$ we have,

$$P(\alpha) = P(\tilde{\alpha})\exp\left(-\frac{1}{2}\lambda(\alpha - \tilde{\alpha})^2\right) \tag{43}$$

where,

$$\lambda \equiv -\frac{\partial^2 \ln \Omega}{\partial \alpha_k^2} = -\frac{\partial \beta}{\partial \alpha_k} \tag{44}$$

Going back to the path integral of equation (27), we identify $P(\alpha)$ with $\hat{\Psi}\{\alpha_k\}$, and as a result we have for the distribution of quantum realities in each segment $k$ the following form:

$$\hat{\Psi}(\alpha_k) = \hat{\Psi}(\tilde{\alpha})\exp\left(-\frac{1}{2}\lambda(\alpha_k - \tilde{\alpha})^2\right) \tag{45}$$

Whereby $\hat{\Psi}\{\alpha_k\}$ is suitably normalized, such that the integration over $\alpha_k$ will be equal to unity. In order to gauge better the orders of magnitude of this distribution, we need to put a meaning into the scaling constant $\lambda$. Another important result that is derived by the general statistics of large ensembles ([11] 2.5) is the fact that for a system with average energy that is sufficiently removed from the ground state, the rate with which the number of states grows as a function of energy is given according to,

$$\Omega \propto \alpha^f \tag{46}$$

where $f$ is the number of degrees of freedom in the system and is determined by some integer constant. This, therefore means that for our case given that the average system energy is taken at $\tilde{\alpha}$, we have:

$$\beta = \left.\frac{\partial \ln \Omega}{\partial \alpha}\right|_{\tilde{\alpha}} = \frac{f}{\tilde{\alpha}} \tag{47}$$

$$\lambda = -\left.\frac{\partial^2 \ln \Omega}{\partial \alpha^2}\right|_{\tilde{\alpha}} = \frac{f}{\tilde{\alpha}^2} > 0 \tag{48}$$

Now, typically for a system characterized by a very large ensemble, $f$ can be adequately approximated by $N$ the number of particles. In the case of the universons this forces us to make an additional assumption. Namely, if the universons are characterized by a finite non-reducible size, then $f$ will be proportional to that length scale. In addition, one also has to take into account two additional constraints: first, the mass of the moving test particle, in order to eliminate any ambiguities that may arise from mass differences (we do not want different physical behaviors to depend on mass, that will be a violation of the equivalence principle). Second, we would like to ensure a priori that when the scale of motion reaches the planck length, the emergent physics will be that which is identical to the "absolute zero of distance" or DSR physics discussed before. Hence for a particle moving along some path, the number of universon degrees of freedom will be:

$$f = \frac{\delta m l}{l_u} g^{-1}\left(\frac{l}{l_p}\right) \tag{49}$$

where $l$ is the length of the path segment, $\delta m$ is the mass of the test particles, $l_p$ is the planck length, $l_u$ is the size-scale associated with the universons that has dimensions length times mass and is proportional to the planck length, and $g^{-1}$ is defined as follows:

$$g^{-1}\left(\frac{l}{l_p}\right) = \begin{pmatrix} 1 & l \gg l_p \\ 0 & l \approx l_p \end{pmatrix} \tag{50}$$

Now, combining eqns. (48), (49) and (50) inserting into the path integral (eqn (27)), we obtain:

$$\Phi(path) = \int \exp\left(-\frac{\hbar^2}{2}\frac{\delta m \Delta_{0\to 1}}{l_u}\left(\alpha_0 - \frac{1}{\hbar}\right)^2 g^{-1}\left(\frac{\Delta_{0\to 1}}{l_p}\right) + i\alpha_0 S[x_0(c), x_1(c)]\right)\ldots d\alpha_0 \ldots d\alpha_{n-1} d\Delta_{0\to 1} \ldots d\Delta_{n-1\to n} \tag{51}$$

changing variables and integrating over the $\alpha$'s we get,

$$\Phi(path) = \int \exp\left(-\frac{l_u}{2\hbar^2} \sum_k \frac{S[x_k(c), x_{k+1}(c)]^2}{\delta m \Delta_{k \to k+1}} g\left(\frac{\Delta_{k \to k+1}}{l_p}\right) + \frac{i}{\hbar} \sum_k S[x_k(c), x_{k+1}(c)]\right) d\Delta_{0 \to 1} \ldots d\Delta_{n-1 \to n} \qquad (52)$$

$$\Phi(path) = \int \exp\left(-\frac{l_u}{2\hbar^2} \sum_k \frac{S[x_k(c), x_{k+1}(c)]^2}{\delta m (x_{k+1}(c) - x_k)} g\left(\frac{x_{k+1}(c) - x_k}{l_p}\right) + \frac{i}{\hbar} \sum_k S[x_k(c), x_{k+1}(c)]\right) d\Delta_{0 \to 1} \ldots d\Delta_{n-1 \to n} \qquad (53)$$

which is the *vsl*-**mechanics path integral** for a universe described by a universon ideal gas. Note, that this path integral re-derives the Feynman integral with a correction term. One of the key results that is immediately reflected by this form is that if the universon length is 0 the correction term which is linearly dependent on $l_u$ vanishes, and all we have left is the conventional quantum mechanics path integral. Now, if $l_u$ is finite, we will get new physics at a length scale when the *vsl*-mechanics term is not negligible as compared with the quantum mechanics term. In other words, what separates QM from *vsl*-mechanics is the hypothetical existence of the universon length scale $l_u$, and as a result the question that was posed at the beginning of this manuscript can be reposed as – do we live in a universe with non-zero $l_u$ (or one where an "absolute-zero-of-length" scale exists) ?

## VII.   A (RELATIVISTIC) FIELD THEORY VERSION OF *VSL* MECHANICS.

Developing a field theory version of *vsl*-mechanics requires that we adapt the assumptions made for the single particle case to many particles, or the particle density continuum case – through the definition of field. In the following, we will attempt to make the most logical adaptation to the original *vsl*-mechanics assumptions. Note, that in the field version we will take $g^{-1}$ to be unity, as we do not expect to treat the planck length regime in this work. However, the full *vsl*-field path integral must include this term, and its is assumed to be implicitly contained within the field version as well. Therefore, in the *vsl*-mechanics field version we will make the following adaptation:

$$\begin{aligned} x_k &\Rightarrow \eta_k \\ L &\Rightarrow \Gamma \end{aligned} \qquad (54)$$

where $\eta_k$ is now the field density and $\Gamma$ is the Lagrangian density. Therefore, the infinitesimal action is now defined as follows:

$$S \equiv \Gamma \delta = \Gamma\left(\frac{\eta_k - \eta_{k-1}}{l}, \eta_k\right)\delta \qquad (55)$$

where $\delta$ is the 4-volume and $l$ corresponds to a segment from the world path which represents the direction of propagation of the field density.

Similarly, we need to define the universon field degree of freedom parameter. In this case we have a particle distribution, which varies slowly over a segment of world path, and therefore the field "universon degree of freedom" become:

$$\tilde{f}_i = \frac{\Delta n_k}{\varepsilon_u} \tag{56}$$

where $\varepsilon_u$ corresponds to some fundamental universon field energy density, which can then be expressed as.

$$\tilde{f}_i = \frac{\Delta n_k \Delta V}{E_u} \tag{57}$$

where $\Delta V$ is some representative 3-volume, and $E_u$ is some fundamental energy constant which is related to $l_u$, and as we will discuss later to $l_p$ as well.

Now defining the infinitesimal 4-volume as:

$$\delta \equiv \Delta V l \tag{58}$$

we then obtain the following path integral extension using the same arithmetic as was done for the single particle case:

$$\Phi_f(path) = \int \exp\left(\frac{-E_u}{2\hbar^2} \sum_{k,\delta} \frac{\Gamma_k^2 \delta}{\Delta \eta_k / l} + \frac{i}{\hbar} \sum_{k,\delta} \Gamma_k \delta \right) d\eta_0 d\eta_1 \cdots dn_k \tag{59}$$

Now, given that a segment of the world path is defined as:

$$l^2 \equiv \Delta x_\mu n^{\mu\nu} \Delta x_\nu \tag{60}$$

where, in this case $\Delta x_\mu$ is some 4-coordinate, $n^{\mu\nu}$ is the special relativity Minkowski metric, and $\mu\nu$ correspond to the summation convention in the usual fashion, we can then express in the continuum limit the denominator of the path integral *vsl*-mechanics term as follows:

$$\frac{\Delta \eta_k}{l} = \left(\left|\frac{\Delta \eta_k}{l}\right|^2\right)^{\frac{1}{2}} = \left(\left|\frac{\Delta \eta_k}{\Delta x_\alpha} n^{\alpha\beta} \frac{\Delta \eta_k}{\Delta x_\beta}\right|\right)^{\frac{1}{2}} = \left|\partial_\mu \eta\right|_k \tag{61}$$

Finally, we can now obtain the *vsl*-mechanics field-theory Lagrangian density, which can be expressed as follows:

$$\Gamma^{vsl} = \frac{i}{\hbar}\left(1 + i\frac{E_u}{2\hbar}\frac{\Gamma}{|\partial_\mu \eta|}\right)\Gamma \tag{62}$$

Note, that the correction term to the standard field theory lagrangian density is complex. This may be bothersome to some readers, particularly when a classical regime approximation is needed. However, this is a mathematical as well as physical necessity, as the universon length scale is real. Indeed, as we will see in the remaining parts of this work, it is the complex nature of the *vsl* correction term which will enable us to formulate a possible solution to the dark energy, dark matter, and pioneer anomaly problems. A real *vsl* correction term will result in equations of motion that do not match the physical reality. In addition, a complex correction term is necessary from the perspective that the theory needs to reduce to the proper limit in the two extreme cases of QM, and DSR – absolute zero of length physics.

Finally, it is important to note, that universon energy constant $E_u$ has dimensions of length$^4$/sec. Therefore it is tempting to define it as follows:

$$E_u \equiv cV_u \tag{63}$$

where *c* is the conventional speed of light value and $V_n$ is some universon characteristic volume.

## VIII. THE CLASSICAL LIMIT – DO WE LIVE IN A UNIVERSE WHERE $L_U > 0$ ?

Like conventional quantum mechanics, which reduced to classical mechanics in the appropriate classical limit, the changes introduced to the path integral due to the new modification will also generate classical equations of motion. These new equations of motion will be derived in the non-relativistic limit for a simple field characterized by gradient-square energy function, and then applied to a generalized problem in order to compare and contrast the solutions that emerge from the *vsl*-classical equations of motion to the standard classical description. The general feature that will characterize the *vsl*-type equation of motion is the emergence of two conjugate force laws that replace (and reduce to in the appropriate limit) the standard Newtonian single force law description – a direct consequence of the complex *vsl* correction term. The two forces are essentially the eigenforces of the system and correspond to the "observed" inertial response of a moving particle in the *vsl*-universe. That is, the complex *vsl* classical action leads to a radical redefinition of the concept of inertia.

In order to show this, I will first carry out the mathematical minimization procedure, specified by the principle of variation, on some general complex *vsl*-action. We will derive two resultant equations of motion for the field, and it is the solution for these two Siamese twins, in a sense, that minimizes the action. To make this point clearer consider the following general complex action for some arbitrary field:

$$S = \int f(\phi, \partial_\mu \phi) \exp(\pm i g(\phi, \partial_\mu \phi)) d^4 x \tag{64}$$

where $f$ and $g$ may be real or complex functions. The nature of a complex functional is in some sense equivalent to a "vector" concept. In this sense, we expect the minimization of the complex functional to be two separate and independent minimization procedures that are carried out on the real and imaginary parts of these functions respectively. We therefore minimize $S$ in accordance with the standard variational procedure. The complex Euler-Lagrange equation that is derived is then:

$$\partial_\mu \left[ \left( \frac{\delta f}{\delta \partial_\mu \phi} \pm i f \frac{\delta g}{\delta \partial_\mu \phi} \right) \exp(\pm i g) \right] - \left( \frac{\delta f}{\delta \phi} \pm i f \frac{\delta g}{\delta \phi} \right) \exp(\pm i g) = 0 \tag{65}$$

This can be conveniently expressed in matrix form, as follows:

$$\partial_\mu \left[ \begin{pmatrix} \frac{\delta f}{\delta \partial_\mu \phi} \\ f \frac{\delta g}{\delta \partial_\mu \phi} \end{pmatrix} \circ \begin{pmatrix} \cos(g) & -\sin(g) \\ \pm \sin(g) & \pm \cos(g) \end{pmatrix} \right] - \begin{pmatrix} \frac{\delta f}{\delta \phi} \\ f \frac{\delta g}{\delta \phi} \end{pmatrix} \circ \begin{pmatrix} \cos(g) & -\sin(g) \\ \pm \sin(g) & \pm \cos(g) \end{pmatrix} = 0 \tag{66}$$

and as a result can be rewritten as a function of two "inertial forces" as follows:

$$\partial_\mu \begin{pmatrix} F_R \\ F_I \end{pmatrix} = \begin{pmatrix} \rho_R \\ \rho_I \end{pmatrix} \tag{67}$$

thus leading to an extension of Newton's law. This equation means that in a universe where "classical" motion is described via a complex action, two types of inertial responses must result – a "real" or observable component $F_R$, and imaginary or "phase" like component of inertia $F_I$. The concept of inertia can only be completely described by solving both of these coupled equations. From an observational perspective, the modified law of inertia manifests itself by allowing us to observe the two "eigenforces" that correspond to the top and bottom term of the column vector in the left hand side of equation. Therefore, it is the equation of motion described by these two eigenvectors which the observer records, having the understanding that the underlying physics is described by an extended Newton's second law as stipulated by equation (67).

In other words, the physical meaning of a complex action at the classical level is the emergence of two force equations to which a moving particle responds. This is not to say that there are two distinct force fields, but rather two distinct forces that act on a moving particle as a result of a given field. This is in short the meaning of the *vsl/mqr* principle at the classical level. This extension of Newton's second law (and through it the equivalence principle) will naturally require an appropriate redefinition of Galilean and Lorentz invariance to make the latter consistent with the *vsl*-approach. What has been described above is essentially only the "left hand side" of any set of force equations. The right hand side (typically referred to as the "mass" or "source" term) does not trivially emerge from the *vsl/mqr* description at the classical level, and will be derived based on empirical observations in the present work. The proper treatment of the mass term will require an appropriate modification to the equivalence principle in the *vsl*-reality, a point that we shall return to in subsequent sections.

The only case where the mass term can be treated unambiguously is when it is equal to or approaching zero. In the standard Newtonian scheme the only way a force-field can apply a null force is if the force-field itself reduced to zero as a function of the mass. This is not necessarily the case for *vsl*-mechanics. Indeed, a null solution is still possible and desirable, yet a second non-null solution also emerges by virtue of having the field's action on a particle be described by two conjugate forces. The non-null solution is one where "the force-field" converges on a constant value as the mass term slowly reduces to zero. As can be seen by the following generalization, and basing this on dimensional considerations. Let,

$$\Gamma^{vsl} = a(\nabla \phi) + ib(\nabla \phi) \tag{68}$$

since *a* and *b* must have the same dimensions their ratio must be related by some non-dimensional constant. When we obtain the Euler equations for this case, with the mass term converging to 0 we get:

$$\frac{\delta a}{\delta \nabla \phi} + i \frac{\delta b}{\delta \nabla \phi} \to 0 \tag{69}$$

This equation can be satisfied by either a null field, or by a field gradient, which equals a constant. The latter implies that in a *vsl*-universe a free-moving body feels a constant, omni-present force despite not having any pulling masses around. This is a direct consequence of the complex nature of inertia. That is, instead of having the time-tested Newton's second law in response to any force field, in the *vsl/mqr* case a force field not only generates two independent forces, but also brings about a repulsive omni-present force that exists even if the pulling mass applies a negligible tag. Indeed, having a single massive particle as a source for a field in a *vsl*-universe is sufficient to have all observers everywhere measure this force. Another way of looking at this is by reflecting on the fact that a force field, no matter its size, applies this force on moving objects, and as a result – unless the universe is completely free of objects generating the field – this force corresponds to a baseline of accelerated motion. One can therefore look at the existence of this force as evidence for both a nonzero $l_u$ and an absolute zero of distance length scale.

##     IX.    DISCUSSION

In the first of the two *vsl*-mechanics papers, I began by posing the following question: "How would particles behave in a universe that has the exact same properties as ours does except that the speed of light is not constant but local?" By local, I mean that the standard value is measured on a large enough scale, but on ultra-microscopic scales this value can fluctuate. Motion in such a universe can only be described via a probabilistic approach that accounts for all possible sequences of random speed of light value, when one attempts to gauge the path of a particle. Thus, we were led naturally to a path integral formalism in order to describe the motion. The general path integral developed for *vsl*-mechanics can be looked upon as describing a landscape of different possible physics, where each point on this manifold corresponds to a different set of emergent conditions which characterize the motion. The manifold itself is characterized by the underlying distribution of the possible speed of light values that each point in space-time can take. By performing a simple mathematical redefinition, the *vsl* path integral formalism, thus led to an intimate connection between the speed of light parameter $c$ and the planck parameter $\hbar$. In fact, the "physics manifold" can be equivalently viewed as the result of the underlying distribution of the latter. The intimate connection between $c$ and $\hbar$ was already alluded to in other *vsl*-theories (for review see [1]) motivated in part by controversial observations [12, 13] that claim a lack of constancy for the fine structure constant ($\alpha$). In these theories the claim was that if a dimensionless parameter is found not to be constant, one or more of its constituents must then vary. Even though these *vsl*-theories assumed a slow variation in time for the speed of light, the conclusion that such variation may also mean a similar behavior from the planck's constant is valid and similar to the one achieved in the present work.

The *vsl* path-integral formalism, thus, can be looked upon as a generalization of quantum mechanics, which may be viewed as "the trivial" point on this manifold. That is, in order to derive the quantum mechanics path integral, all one has to do is choose a delta function distribution for the planck parameter in eqn. (27) of the accompanying manuscript. From a *vsl*-approach, this translates to an intuitive definition to why motion on the quantum scales appears the way that it does. In a sense, it is an artifact of the assumption that neither $c$ nor $\hbar$ are constants. But quantum mechanics is just one possible solution or reality, as *vsl*-mechanics allows for many more possibilities. In a flat distribution, where all quantum realities are equally probable, we reach the conclusion that displacement measurements on that scale will always result in the same distance value. That is, the distance scale that sets the flat distribution, will be in fact the smallest distance that any measuring apparatus can measure. Smaller distances will simply not make sense. This interpretation is due to the fact that on this special scale the plethora of quantum realities cancel each other's wave-like characteristics, thus resulting in deterministic but completely random distance measurements. This conclusion is one of the fundamental results that emerges from this theory, and is another way of expressing the doubly special relativity concept introduced by [3]. Indeed, the very existence of a minimum distance scale in any context (not just *vsl*) will have profound ramifications on ultra-microscopic physics.

Due to the large amount of possibilities that can be used to define the planck parameter function, it became imperative to design a more physical model for these quantities that will span the two extreme cases of quantum mechanics and the flat distribution described above. I, therefore, assumed that the fabric of the universe is made of "particle-like" objects called universons, whereby each "particle" is characterized by a given set of fundamental parameters (i.e. speed of light, planck's constant, etc.) The universons behave like a thermal gas with a "universon temperature", and the average value of their characteristic parameters measure up to the familiar constants of nature. A point-like particle in motion, in a sense, "sees" varying constants of nature along its path as it moves or tunnels from one universon to the next. In each point in space-time, the same general physics laws apply. That is, special relativity, the various inverse square laws, general relativity, etc. are all defined and hold at every specific position in space-time. Motion to an adjacent coordinate will result in the variation of the parameters, meaning the same general laws will hold, but with different values for the constants defining them. I believe that a *vsl* universe is still relativistic in the sense that there are no inertial reference frame, and that a certain type of generalized Lorentz invariance symmetry may hold. Moreover, it is tempting to speculate that the *vsl* concept can be an additional group of symmetries that together with Lorentz group forms a larger and all-encompassing symmetry group. However, I concede that the point of inertial frames and Lorentz invariance is controversial in the present *vsl* context, and will require a detailed treatment in a fully relativistic version of *vsl*-mechanics, which is yet to be developed.

Another important implication of the *vsl*-approach is a new insight into the relationship between quantum and classical regimes in the laws of physics. In the present context, the classical equations of motion that emerge through a familiar Euler-Lagrange operation do not resemble at all their quantum counterparts. Indeed, the prediction here is that a *vsl/mqr*-assumption at the ultramicroscopic level transposes into a radical redefinition of inertia at the macroscopic level. This is one of the more important results of the present work, and provides an interesting connection between the concept of inertia and microscopic definitions of motion. Indeed, the accompanying manuscript is dedicated to exploring this aspect of vsl-mechanics, in applying it to the Newtonian gravitational field.